\documentclass[prb,twocolumn,showpacs]{revtex4}
\usepackage{graphicx}
\usepackage{amsmath}
\usepackage{amssymb}

\begin{document}


\title{Anomalous Absorption Line in the Magneto-Optical Response of Graphene}

\author{V.P.~Gusynin$^{1}$}
\author{S.G.~Sharapov$^{2}$}
\author{J.P.~Carbotte$^{2}$}

\affiliation{$^1$ Bogolyubov Institute for Theoretical Physics,
        14-b Metrologicheskaya Street, Kiev, 03143, Ukraine\\
        $^2$ Department of Physics and Astronomy, McMaster University,
        Hamilton, Ontario, Canada, L8S 4M1}

\date{\today }

\begin{abstract}
The intensity as well as position in energy of the absorption lines
in the infrared conductivity of graphene, both exhibit features that
are directly related to the Dirac nature of its quasiparticles. We
show that the evolution of the pattern of absorption lines as the
chemical potential is varied encodes the information about the
presence of the anomalous lowest Landau level. The first absorption
line related to this level always appears with full intensity or is
entirely missing, while  all other lines disappear in two steps. We
demonstrate that if a gap develops, the main absorption line splits
into two provided that the chemical potential is greater than or
equal to the gap.
\end{abstract}




\maketitle

While graphene has only recently been isolated
\cite{Novoselov2004Science}, several anomalous dc properties already
appear well established
\cite{Geim2005Nature,Kim2005Nature,Zhang2006PRL}. This includes an
unconventional integer quantum Hall effect with filling factor $\nu
= \pm 2, \pm 6, \pm 10,\ldots$
\cite{Geim2005Nature,Kim2005Nature,Zheng2002PRB,Gusynin2005PRL,Peres2006PRB}
which changes to $\nu=0, \pm1,\pm2,\pm4, \ldots$ at a magnetic field
$B$ above $20 \, \mbox{T}$ Ref.~\cite{Zhang2006PRL}, which may
indicate an opening of a gap. Also observed is a Berry phase of
$\pi$ \cite{Geim2005Nature,Kim2005Nature,Gusynin2005PRL} which
increases to $2\pi$ in bilayers \cite{Novoselov2006NaturePh}.
These spectacular dc properties can be traced to the nature of the
quasiparticles in graphene which are massless Dirac fermions
governed by $2+1$ dimensional quantum electrodynamics
\cite{Semenoff1984PRL,Haldane1988PRL}. Optical conductivity
measurements play a central role in solid state physics and have
provided additional information on the dynamics of charge carriers
and its change under external perturbations, not directly available
in dc measurements. Here we present a theory of the diagonal optical
conductivity of Dirac quasiparticles in an external magnetic field
and highlight several unusual features, not present in a
conventional two dimensional electron gas, which bear directly on
their unconventional dynamics.  We also investigate an emerging new
physics of graphene related to the sublattice symmetry breaking
reported in Ref.~\cite{Zhang2006PRL}.

In the presence of a magnetic field, Landau levels (LLs) form in the
electronic density of states and transitions between these levels
give rise to absorption lines in the optical conductivity. In this
letter we find that the selection rules, which set the frequencies
of the lines and their relative spectral weight, provide a distinct
signature of the presence of the Dirac quasiparticles. Not only do
the positions and the intensities of the absorption lines scale with
$\sqrt{B}$ rather than $B$, but the line corresponding to the
transition from the lowest, $n=0$ to the $n=1$ (Landau level) LL is
anomalous due to the special Dirac character of the $n=0$ level.
This level and no others can participate in both inter and intraband
transitions. As a consequence, the first absorption line always
appears with full intensity or is entirely missing, while all other
lines can also be seen to have in addition half intensity depending
on the position of chemical potential $\mu$ relative to the LL
energies. The chemical potential $\mu$ is tunable in experiments by
application of a gate bias voltage to a field effect device
\cite{Novoselov2004Science,Geim2005Nature,Kim2005Nature,Zhang2006PRL}.

Expressions for the absorptive part of the diagonal as well as Hall
conductivity on which our calculations are based can be found in
\cite{Gusynin2006PRB,Gusynin2006PRL,Peres2006PRB}. In the limit of
zero impurity scattering rate, $\Gamma \to 0$, the diagonal
conductivity as a function of the photon energy $\Omega$ is
\begin{equation}
\begin{split}
\label{sigma-gap} & \mbox{Re} \, \sigma_{xx}(\Omega) = \frac{e^2}{h}
\frac{2 v_F^2 |eB|
\hbar}{c \Omega} \frac{\pi}{2} \times \\
&\sum_{n=0}^{\infty} \left[ \left(1+\frac{\Delta^2}{M_n
M_{n+1}}\right) A(T) \delta(M_n + M_{n+1} - \Omega) \right.\\
& + \left. \left(1-\frac{\Delta^2}{M_n M_{n+1}}\right) B(T)
\delta(M_n - M_{n+1} + \Omega)  \right],
\end{split}
\end{equation}
where the thermal factors are $A(T) =
n_F(-M_{n+1})-n_F(M_n)+n_F(-M_n)-n_F(M_{n+1})$ and $B(T) = n_F(M_n)
- n_F(M_{n+1}) + n_F(-M_{n+1}) - n_F(-M_n)$  with the Fermi
distribution $n_F(\omega) = [\exp[(\omega-\mu)/T] +1]^{-1}$ (we set
the Boltzmann  constant $k_B=1$). To include damping, the $n$'th
delta function in Eq.~(\ref{sigma-gap}) is replaced by a Lorentzian
of width $\Gamma_n + \Gamma_{n+1}$ with $\Gamma_n$ the width of
$n$'th level. Also in Eq.~(\ref{sigma-gap}) a term $\Omega \to
-\Omega$ is to be subtracted. In the presence of an excitonic gap
$\Delta$ \cite{Khveshchenko2001PRL,Gorbar2002PRB} the LL energies
are $E_n = \pm M_n$, with $M_n = \sqrt{\Delta^2+ 2n |eB| \hbar
v_F^2/c}$, and $v_F$ is the Fermi velocity in graphene, $\hbar$ is
Planck's constant $h$ over $2\pi$, $c$ is the velocity of light and
$e$ is the charge of the electron. The field $B$ is applied
perpendicular to the plane and for $B=1\,\mbox{T}$ the energy scale
$M_1\approx 420\, \mbox{K}$ for $v_F \approx 10^6\, \mbox{m/s}$ and
$\Delta=0$. As for the conventional case, the selection rules allow
transitions only between adjacent LLs (see
Ref.~\cite{Gusynin2006PRB} for details).
\begin{figure}
\centering{
\includegraphics[width=7cm]{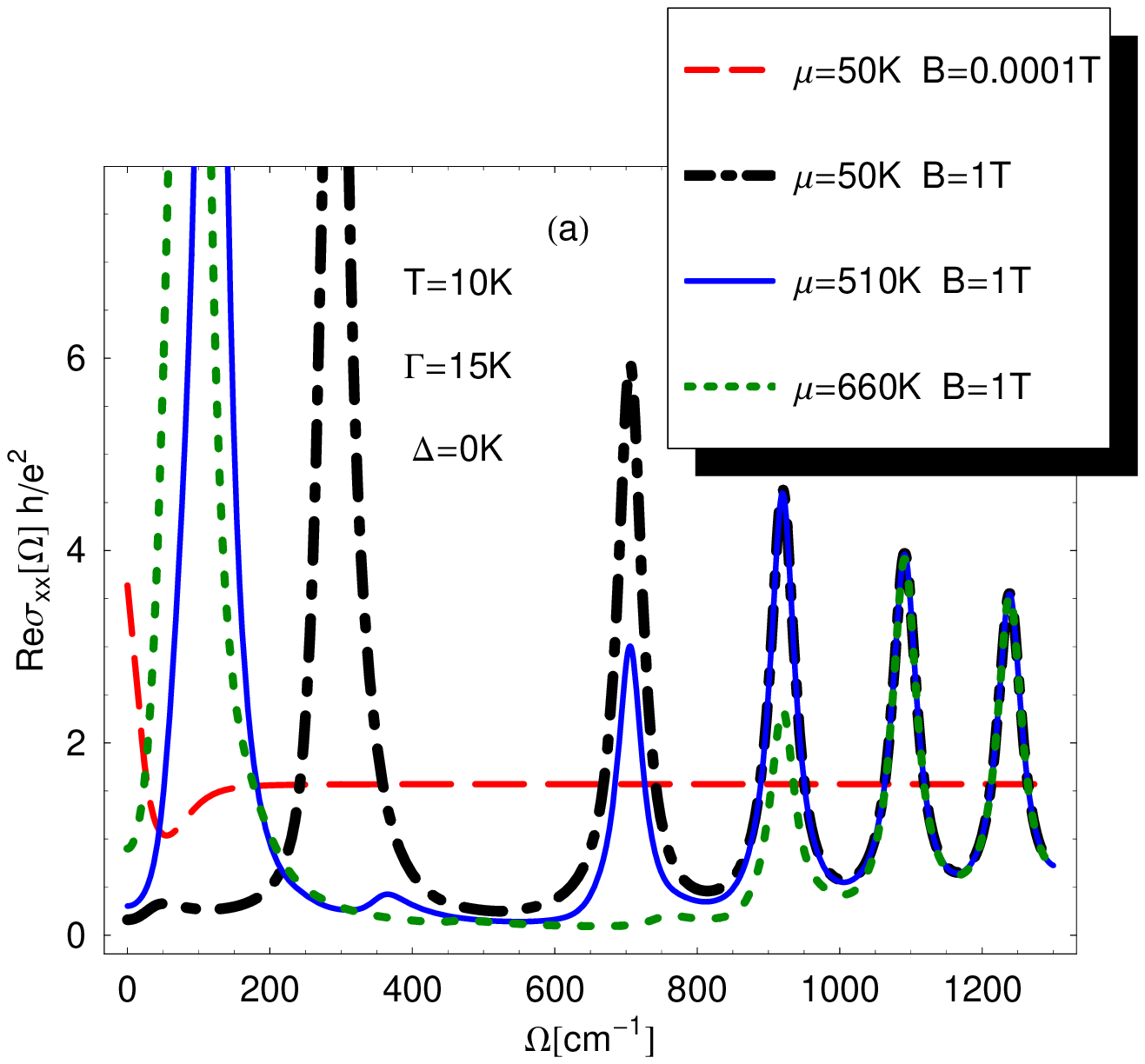}}
\centering{
\includegraphics[width=6.5cm]{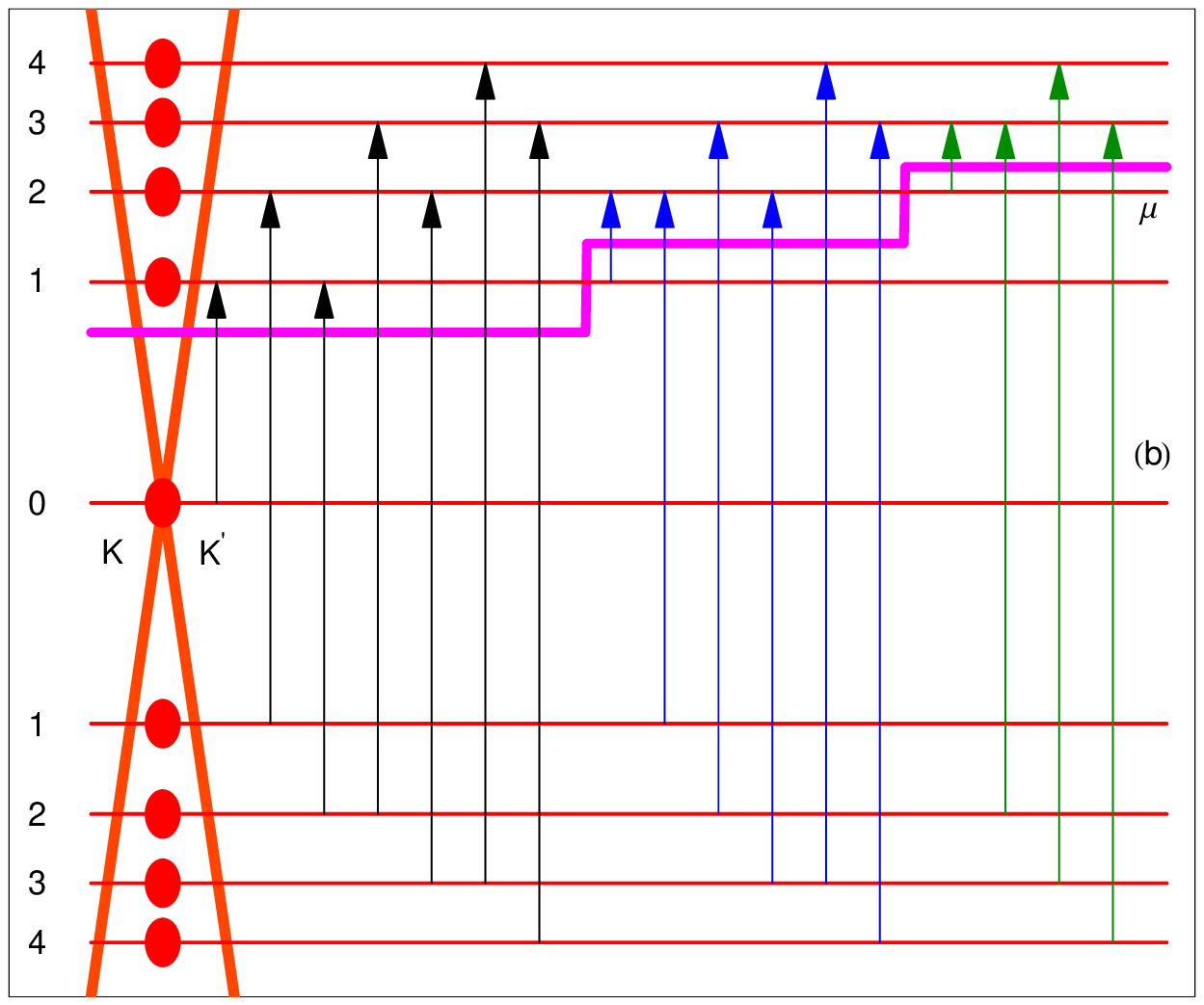}}
\caption{The absorption peaks and  corresponding transitions. (a)
Real part of the conductivity, $\mbox{Re} \, \sigma_{xx}(\Omega)$ in
units of $e^2/h$ vs frequency $\Omega$ in $\mbox{cm}^{-1}$ for
temperature $T=10 \mbox{K}$ and scattering rate $\Gamma =15
\mbox{K}$. Long dashed (red), the chemical potential $\mu = 50
\mbox{K}$ and the magnetic field $B=10^{-4} \mbox{T}$, dash-dotted
(black) $\mu = 50 \mbox{K}$, solid (blue) $\mu = 510 \mbox{K}$,
short dashed (green) $\mu = 660 \mbox{K}$ all three for
$B=1\mbox{T}$. (b) A schematic of the allowed transitions between
LLs $n=0,1,\ldots 4$ shown as heavy (red) dots for both positive and
negative Dirac cones. The pair of cones at points $\mathbf{K}$ and
$\mathbf{K}^\prime$ in the Brillouin zone (see Fig.~3 (b)) are
combined. Three values of chemical potential are shown as the
(violet) heavy solid line. At the left, $\mu$ falls between $M_0$
and $M_1$; middle, between $M_1$ and $M_2$; and on the right,
between $M_2$ and $M_3$. The first line on the left which
corresponds to the first peak in dash-dotted line in upper panel is
unusual in that it arises from both interband and intraband
transitions and ceases to occur for all other values of chemical
potential, while all other interband lines first drop to half their
intensity before disappearing entirely as $\mu$ crosses to higher
energy levels. } \label{fig:1}
\end{figure}

Our numerical results for  $\mbox{Re} \, \sigma_{xx}(\Omega)$  are
given in Fig.~\ref{fig:1}~(a), where it is plotted in units $e^2/h$
as a function of $\Omega$ for $\Delta=0$. The same constant
electronic scattering rate $\Gamma = 15 \mbox{K}$
\cite{Zhang2006PRL} was used for all LLs, an assumption that could
easily be relaxed. Except for the long dashed (red) curve for which
the magnetic field $B=10^{-4} \mbox{T}$, and is shown only for
comparison, $B=1 \mbox{T}$ for other curves. They differ only in
value of  chemical potential $\mu$. The dash-dotted (black) curve
has $M_0<\mu <M_1$; solid (blue) curve has $M_1 < \mu < M_2$, and
the short dashed (green) curve has $M_2 < \mu <M_3$. The curves
remain essentially unchanged for any values of $\mu$ between $M_{N}$
and $M_{N+1}$ and evolve rapidly from one pattern to the next as a
new LL is crossed, for a change in chemical potential of order $T$
or $\Gamma$ whichever is greater.

Referring to Fig.~\ref{fig:1}(a), in the dash-dotted (black) curve
the lowest energy peak is at $M_1$ and the others at $M_1+M_2$,
$M_2+M_3, \ldots$ all correspond to interband transitions from
negative to positive Dirac cones. As $\mu$ is increased to cross the
$n=1$ LL [solid (blue) curve], the first peak in the dash-dotted
(black) curve disappears and the intensity of the next higher peak
drops to half its value, while all other peaks remain the same. In
addition, a new intraband peak appears at $M_2 - M_1$ and its
intensity picks up the lost optical spectral weight from the missing
and/or reduced interband transitions. Comparing with the
short-dashed (green) curve, the first two interband lines have
disappeared,  the third has halved and the intraband transition is
at $M_3 - M_2 $. The first line of the interband series is different
in that it never appears with half intensity, while the others do.
It is the only one which can be considered as involving both
interband and intraband transitions as can be seen in
Fig.~\ref{fig:1}~(b). The energy level schematic shown in this
figure summarizes our results. It shows how the first interband line
on the far left appears alone, while all others come in pairs. This
transition exists only for values of the chemical potential less
than $M_1$, while the others first drop to half intensity as another
LL is crossed before disappearing at the next crossing.
\begin{figure}[h]
\centering{
\includegraphics[width=8cm]{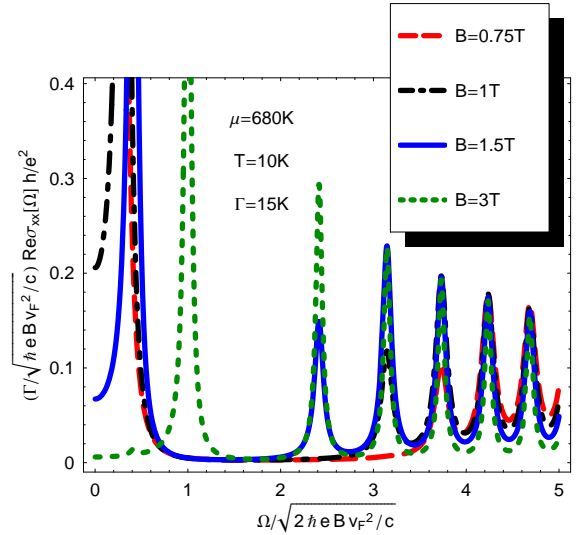}}
\caption{Real part of the longitudinal conductivity, $\mbox{Re} \,
\sigma_{xx}(\Omega)$ in units of $e^2/h$ multiplied by
$\Gamma/\sqrt{\hbar eB v_F^2/c}$ as a function of normalized photon
energy $\Omega/\sqrt{2\hbar eB v_F^2/c}$. The temperature is $T=10
\mbox{K}$, the scattering rate $\Gamma=15 \mbox{K}$ and the chemical
potential $\mu=680 \mbox{K}$. The long dashed (red) curve is for
$B=0.75 \mbox{T}$, dash-dotted (black) for $B=1 \mbox{T}$, solid
(blue) for $B=1.5 \mbox{T}$ and short-dashed (green) for $B= 3
\mbox{T}$.} \label{fig:2}
\end{figure}
To verify these predictions optical experiments on graphene based
field effect device would allow one to change the chemical potential
as in Fig.~\ref{fig:1} through a change in gate voltage. Such
optical experiments have been reported by Li {\it el al.\/}
\cite{Basov-organic} in organic metals. It is however also possible
to get the same information for a fixed chemical potential if
$\mbox{Re} \, \sigma_{xx}(\Omega)$ is measured at a few well chosen
values of the external magnetic field $B$. This is illustrated in
Fig.~\ref{fig:2}. To see the desired effect it was convenient to
normalize $\Omega$ to $\sqrt{2\hbar eB v_F^2/c}$ and to multiply the
vertical scale by the dimensionless quantity $\Gamma/\sqrt{\hbar eB
v_F^2/c}$. This arrangement guarantees that the various interband
lines will all fall at the same normalized value of
$\Omega/\sqrt{2\hbar eB v_F^2/c}$ and the intensity of the various
lines follow the same pattern as in Fig.~\ref{fig:1}.

To understand the intensity that is to be assigned to the various
lines and their changes with $\mu$ it is sufficient to take the
limit $T \to 0$ in Eq.~(\ref{sigma-gap}).
For chemical potential $\mu \in [M_N,M_{N+1}]$ we find ($\Omega >0$)
\begin{equation}
\begin{split}
\label{Re-sigma_xx} \mbox{Re} \, \sigma_{xx}(\Omega)&  =
\frac{e^2}{h}
M_1 \frac{\pi}{2}\\
& \sum_{n=0}^{\infty}  \left\{\left[
                             \begin{array}{ccc}
                               0 & \mbox{for} & n < N \\
                               1 & \mbox{for} & n=N \\
                               2 & \mbox{for} & n>N  \\
                             \end{array}
                           \right] \frac{\delta(M_n+M_{n+1} -
                           \Omega)}{\sqrt{n+1}+\sqrt{n}}\right.\\
& \left. + \left[
             \begin{array}{ccc}
               1 & \mbox{for} & n=N \\
               0 & \mbox{for} & n\neq N \\
             \end{array}
           \right] \frac{\delta(M_n -M_{n+1} + \Omega)}{\sqrt{n+1}-\sqrt{n}}
\right\},
\end{split}
\end{equation}
where the first and second terms give the interband and intraband
transitions, respectively, at the energies indicated by the
$\delta$-functions. The denominators come from the $1/\Omega$ factor
in Eq.~(\ref{sigma-gap}). In units of $(e^2/h) M_1 (\pi/2)$ which
goes like $\sqrt{B}$, the optical spectral weight of various lines
is given by the quantities multiplying $\delta$-functions in
Eq.~(\ref{Re-sigma_xx}). For $N>0$ all interband lines below $n=N$
are missing and the intensity of the line at $n=N$ is half of
$2/(\sqrt{n+1}+\sqrt{n})$. The total optical spectral weight lost is
\begin{equation}
\sum_{n=0}^{N-1} \frac{2}{\sqrt{n+1}+\sqrt{n}} +
\frac{1}{\sqrt{N+1}+\sqrt{N}} = \frac{1}{\sqrt{N+1}-\sqrt{N}}.
\end{equation}
The right hand side is exactly the intensity of the corresponding
intraband line given by the second term in (\ref{Re-sigma_xx}). We
verified that the integrated optical spectral weight under
$\mbox{Re} \, \sigma_{xx}(\Omega)$ up to energy $\Omega_{max}$ is
linear in $\Omega_{max}$ and approximately equal to $(e^2/h) (\pi/2)
\Omega_{max}$ provided $\Omega_{max}$ is taken large enough for many
LLs to be involved. This linear in $\Omega_{max}$ behavior which
holds as well for $B=0$ is another signature of the existence of the
Dirac cones. In ordinary metals for $B=0$ the optical sum would
rapidly saturate to a constant value as $\Omega_{max}$ is increased
beyond the  scattering rate $\Gamma$. The above rules for the
intensity as well as the peculiar behavior of the first inter-
(intra-) band line as $\mu$ is increased from zero are specific to
graphene and as yet have not been verified experimentally, although
some data exist as we now describe.
\begin{figure}
\centering{
\includegraphics[width=7cm]{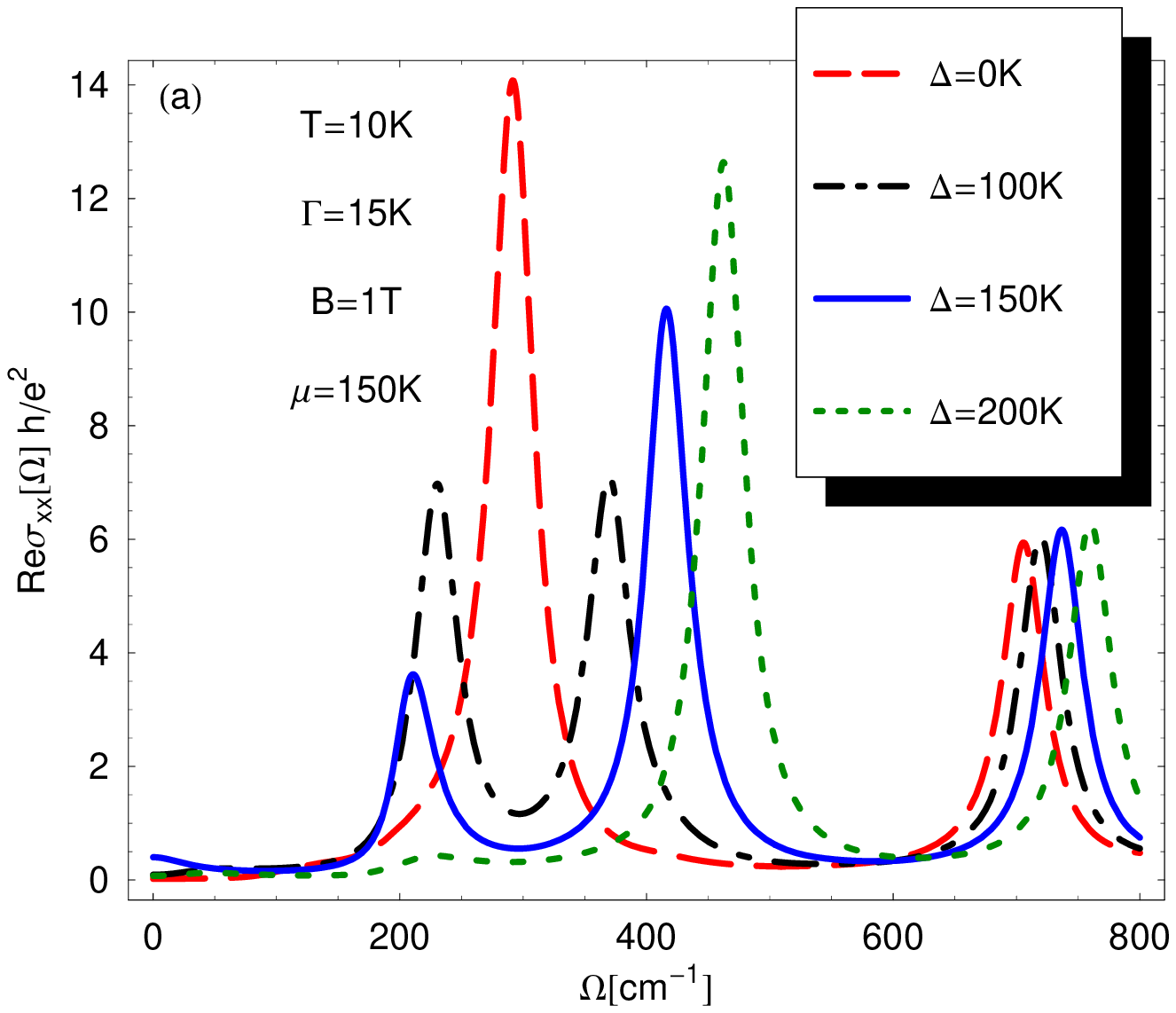}}
\centering{
\includegraphics[width=6.5cm]{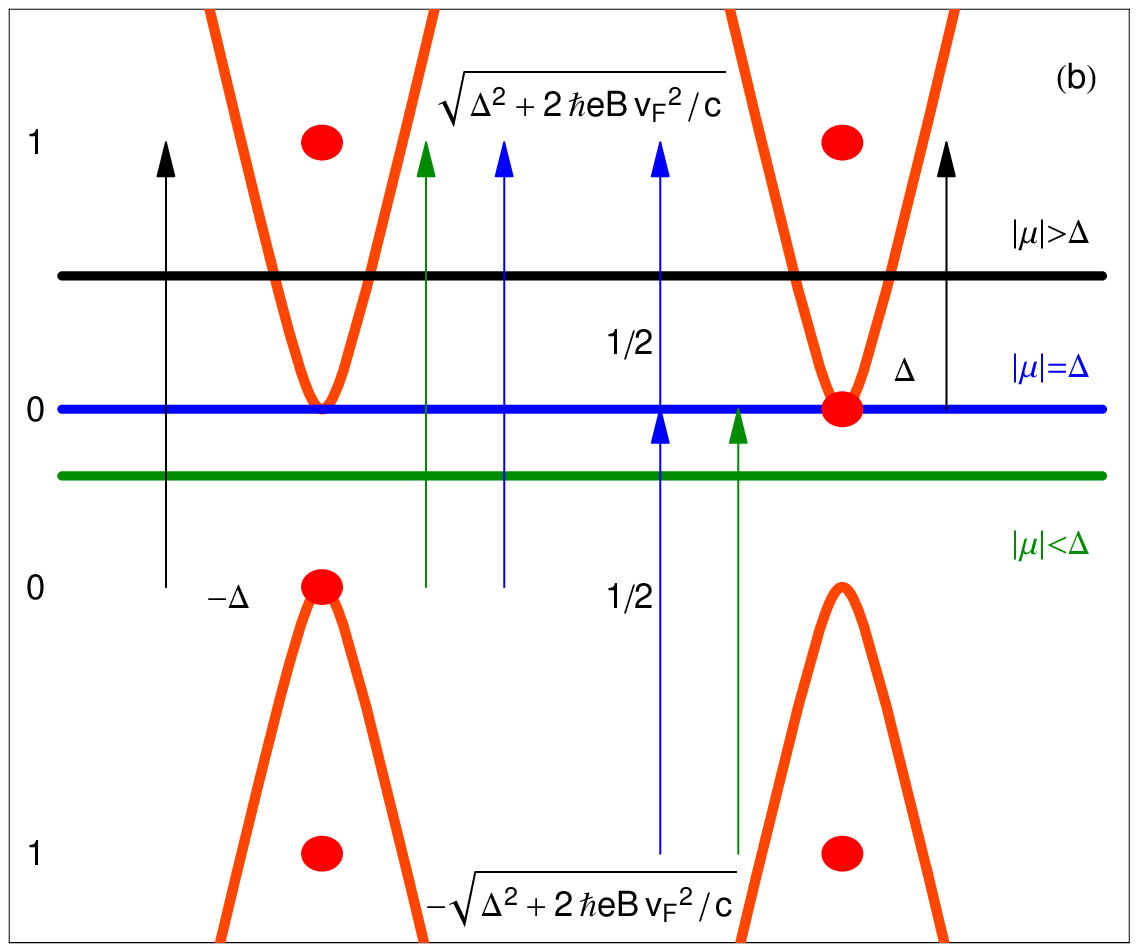}
} \caption{The absorption peaks and the corresponding transitions
when the lowest LL is split. (a) Real part of the longitudinal
conductivity, $\mbox{Re} \, \sigma_{xx}(\Omega)$ in units of $e^2/h$
vs frequency $\Omega$ in $\mbox{cm}^{-1}$ for temperature $T=10
\mbox{K}$, $\Gamma =15 \mbox{K}$, $B=1 \mbox{T}$ and chemical
potential $\mu = 150 \mbox{K}$ for 4 values of the excitonic gap
$\Delta$. Long dashed (red) $\Delta = 0 \mbox{K}$, dash-dotted
(black) $\Delta = 100 \mbox{K}$, solid (blue) $\Delta = 150
\mbox{K}$, short dashed (green) $\Delta = 200 \mbox{K}$. (b)
Possible optical transitions between LLs with an excitonic gap
$\Delta$. Two pairs of cones at $\mathbf{K}$ and $\mathbf{K}^\prime$
points in the Brillouin zone are drawn. Only LLs with $n=0$ with
energies $-\Delta$ (at $\mathbf{K}$ point) and $\Delta$ (at
$\mathbf{K}^\prime$ point), and $n=1$ with energies $\pm M_1$,
respectively, are shown. Three values of chemical potential are
considered $|\mu| < \Delta$ (green), $\mu = \Delta$ (blue) and
$|\mu| > \Delta$ (black). For the case $\mu = \Delta$ the state at
energy $E_0= \Delta$ is both occupied and unoccupied with
probability $1/2$.} \label{fig:3}
\end{figure}
The early magneto-reflectance data \cite{Toy1977PRB} in fields of
order of $1$ to $10 \, \mbox{T}$ in graphite assigned LL lines to
the $H$-point in the Brillouin zone. Some of these have central
frequencies which vary as the square root \cite{Basov-foot} of $B$
as expected in graphene. More recent data by Li {\em et al.}
\cite{Li2006} who use high fields up to $20\, \mbox{T}$, show
conventional linear in $B$ behavior. However, very recently,
infrared transmission data \cite{Sadowski2006} on ultrathin
epitaxial graphite samples \cite{Berger2006Science} in fields up to
$4 \,\mbox{T}$ have become available and provide $\mbox{Re} \,
\sigma_{xx}(\Omega)$. A $\sqrt{B}$ scaling for both the energy and
the intensity of the first line was taken as evidence for Dirac
quasiparticles. In view of the known sensitivity of some properties
to coupling between layers it is not obvious that the various carbon
sheets in the samples are sufficiently decoupled to be treated as
isolated graphene sheets \cite{Basov-foot}. Further, to understand
the results  it is necessary to assume that there is a variation in
charging, i.e. each sheet can have different chemical potential. We
have analyzed the intensities of the one intraband and of the three
interband transitions reported in \cite{Sadowski2006}. The intensity
of the first interband line (B) in Fig.~1 of
Ref.~\cite{Sadowski2006}, which can occur only for $|\mu| < M_1$, is
the largest and that of the intraband line (A) is smaller by a
factor of $0.36$. This can only be understood if there are at least
three times more sheets for which the chemical potential falls below
$M_1$ than above. Further, the ratio of the intensity of the second
(C) and the third (D) interband lines to the first (B) are $0.34$
and $0.19$, respectively. These values are somewhat smaller than
$0.41$ and $0.32$ expected from our theory, coming from the sheets
with $|\mu|< M_1$. Adding in  the other sheets with $|\mu| > M_1 $
only increase the above discrepancy, because they  add to the second
and third interband lines, but not to the first. While the above
provides some indirect support of our theory, further experiments
are needed to verify the main features described here. A case for
further studies is a bilayer \cite{Abergel2006} where it is known
\cite{Novoselov2006NaturePh} that the two lowest LLs both lie at
zero energy and this should change the pattern of behavior of the
absorption lines described here.

The ac conductivity may also be used to get information on the
opening of a possible gap related to the sublattice symmetry
breaking in graphene
\cite{Zhang2006PRL,Khveshchenko2001PRL,Gorbar2002PRB,Nomura2006,Alicea2006,Gusynin2006catalysis}.
In Fig.~\ref{fig:3}~(a) we show results for $\mbox{Re} \,
\sigma_{xx}(\Omega)$ plotted using the generalization of
Eq.~(\ref{sigma-gap}) to finite $\Gamma$. Four values of the
excitonic gap $\Delta$ are considered. For $\Delta < |\mu|$ the
single peak at $294\, \mbox{cm}^{-1}$ has split into two of nearly
equal intensity, for $|\mu| = \Delta$ the lower peak has
approximately $1/3$ the intensity of the upper peak and for $\Delta
> |\mu|$ a single peak is seen at the higher energy.
Fig.~\ref{fig:3}~(b) gives a schematic of the levels involved and
shows the transitions. For convenience we have fixed the gap in the
figure and considered three values of chemical potential. For $|\mu|
< \Delta$ the interband transition $-M_0$ to $M_1$ (LHS cone) and
$-M_1$ to $M_0$ (RHS cone) are both possible and have the same
energy $M_1+\Delta$, so only one line is seen shifted to higher
energy. For $|\mu| = \Delta$ one needs to account for the fact that
the state at $n=0$ (RHS cone) is occupied (unoccupied) with
probability $1/2$. The possible interband transitions are $-M_0$  to
$M_1$ (LHS cone) with energy $M_1 + \Delta$ and $-M_1$ to $M_0$ with
energy $M_1 + \Delta$, and an intraband transition of $M_0$ to $M_1$
with energy $M_1 - \Delta$ (RHS cone) each weighted by $1/2$ rather
than by $1$. This gives a $3$ to $1$ ratio favoring the higher
energy line. Finally, for $|\mu| > \Delta$, the $-M_0$ to $M_1$
interband transition (LHS) has energy $M_1 +\Delta$ and the
intraband transition of $M_0$ to $M_1$ (RHS) has energy $M_1 -
\Delta$. The two lines have nearly the same intensity. It is clear
that optics provides a useful probe for observing various kinds of
gaps and here the excitonic gap is taken as an example.

In summary, we have found an anomaly in the first absorption peak in
the optical conductivity of graphene. When the chemical potential,
tunable by the gate voltage, sweeps through the various LL energies,
this peak never halves its intensity before disappearing while the
other lines do. This pattern of behavior is specific to the Dirac
nature of quasiparticles in graphene and should be easily observable
in the  experiments. It originates from the special structure of LLs
in graphene, where each positive energy level has its negative
counterpart, except for the zero energy lowest LL. For the Shubnikov
de Haas oscillations \cite{Gusynin2005PRB} this structure of LLs
resulted in a phase shift of $\pi$ used to identify these
quasiparticles in the dc measurements
\cite{Geim2005Nature,Kim2005Nature}. Furthermore, should an
excitonic gap open in high magnetic field, its presence will lead to
a specific signature in an ac optical experiment. For $\Delta <
|\mu| < M_1$, the $n=0$ to $n=1$ transition will split into two
lines of nearly equal intensity, while for $|\mu| = \Delta$ the
intensity ratio of higher to lower energy line will be about $3$.
For $|\mu| < \Delta$ the line will simply shift upward without
splitting.

The work of V.P.G. was supported by the SCOPES-project IB7320-110848
of the Swiss NSF. Work by J.P.C. and S.G.Sh. was supported by NSERC
and CIAR. We thank D.N.~Basov, A.K.~Geim and J.S.~Hwang for
discussions and E.J.~Nicol for clarifying discussions and
suggestions on the manuscript.

\end{document}